\documentclass[journal=jpccck,manuscript=article]{achemso}
\usepackage[version=3]{mhchem}
\usepackage{subfigure}
\usepackage{braket}
\author{G.L.~Celardo}
\affiliation{Interdisciplinary Laboratories for Advanced Materials Physics, Universit\`a Cattolica del Sacro Cuore, via Musei 41, I-25121 Brescia, Italy}
\alsoaffiliation{Dipartimento di Matematica e Fisica, Universit\`a Cattolica del Sacro Cuore, via Musei 41, I-25121 Brescia, Italy}
\alsoaffiliation{Istituto Nazionale di Fisica Nucleare,  Sez. di Pavia, via Bassi 6, I-27100,  Pavia, Italy}
\email{nicedirac@gmail.com}
\author{D.~Archetti}
\affiliation{Dipartimento di Matematica e Fisica, Universit\`a Cattolica del Sacro Cuore, via Musei 41, I-25121 Brescia, Italy}
\author{G.~Ferrini}
\affiliation{Interdisciplinary Laboratories for Advanced Materials Physics, Universit\`a Cattolica del Sacro Cuore, via Musei 41, I-25121 Brescia, Italy}
\alsoaffiliation{Dipartimento di Matematica e Fisica, Universit\`a Cattolica del Sacro Cuore, via Musei 41, I-25121 Brescia, Italy}
\author{L.~Gavioli}
\affiliation{Interdisciplinary Laboratories for Advanced Materials Physics, Universit\`a Cattolica del Sacro Cuore, via Musei 41, I-25121 Brescia, Italy}
\alsoaffiliation{Dipartimento di Matematica e Fisica, Universit\`a
  Cattolica del Sacro Cuore, via Musei 41, I-25121 Brescia, Italy}
\author{P.~Pingue}
\affiliation{Laboratorio NEST - Scuola Normale Superiore, and Istituto Nanoscienze - CNR, Piazza San Silvestro 12, I-56127 Pisa, Italy}
\author{E.~Cavaliere}
\affiliation{Interdisciplinary Laboratories for Advanced Materials Physics, Universit\`a Cattolica del Sacro Cuore, via Musei 41, I-25121 Brescia, Italy}
\alsoaffiliation{Dipartimento di Matematica e Fisica, Universit\`a Cattolica del Sacro Cuore, via Musei 41, I-25121 Brescia, Italy}

\title{Evidence of diffusive fractal aggregation of $TiO_2$ nanoparticles by femtosecond laser ablation at ambient conditions}

\keywords{Pulsed Laser Deposition, Fractal structures, Diffusion and Aggregation}
\begin{document}

\begin{tocentry}

\includegraphics[scale=0.4]{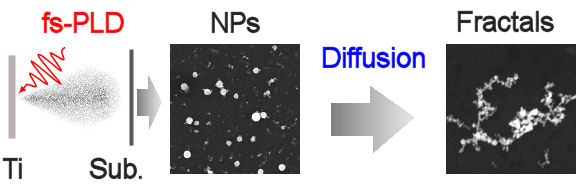}

\end{tocentry}

\begin{abstract}
The specific mechanisms which leads to the formation of fractal nanostructures by pulsed laser deposition remain elusive despite intense research efforts, motivated mainly by the technological interest in obtaining tailored nanostructures with simple and scalable production methods.
Here we focus on fractal nanostructures of titanium dioxide, $TiO_2$, a strategic material for many applications, obtained by femtosecond laser ablation at ambient conditions. 
We model the fractal formation through extensive Monte Carlo simulations based on a set of minimal assumptions: irreversible sticking and size independent diffusion. 
Our model is able to reproduce the fractal dimensions and the area distributions of the nanostructures obtained in the experiments for different densities of the ablated material.
The comparison of theory and experiment show that such fractal aggregates are formed after landing of the ablated material on the substrate surface by a diffusive mechanism. 
Finally we discuss the role of the thermal conductivity of the substrate and  the laser fluence on the properties of the fractal nanostructures.
Our results represent an advancement towards controlling the production of fractal nanostructures by pulsed laser deposition.
\end{abstract}


\section{ Introduction.}
Fractal structures are commonly found in different natural processes~\cite{Fractal}.
While their geometrical features are quite universal, the mechanisms of their formation can be very different. 
In view of the technological applications, fractal nanostructures at the nanoscale have been widely  investigated in surface science.
They are at the base of recent proposals for sensing devices~\cite{acs1} and optical devices~\cite{acs2}. 
The realization of fractal nanostructures would also allow enhancing the selectivity behavior of catalyst material with high porosity~\cite{cat} or to increase the performances of super-capacitors~\cite{tech8}.
Understanding the mechanism of fractal nanostructures formation with the aid of theoretical models is an essential step towards building engineered microdevices with tailored properties.  

Fractal nanostructures have been obtained on various substrates, by deposition  
of the target material ejected as a plume due to the laser ablation process~\cite{nanof1,nanof2}.
This technique, known as pulsed laser deposition (PLD), has been used in various environments (liquids, high gas pressures) and with different laser pulses characteristics (pulse length, wavelength, fluence, polarization).
Fractal aggregates obtained from thousands of ns pulses have been demonstrated in water~\cite{28} and at high argon pressure ~\cite{24,water}.
When the laser pulse duration is reduced to less than a picosecond, PLD enters in the femtosecond regime (fs-PLD), and a different physical mechanism for ablation sets in~\cite{np3}.
In this case, since the pulse duration is shorter than the electron-phonon relaxation time, which is of the order of few picoseconds, the ablation mechanism is not due to thermal melting as in the ns case.
Fractal nanostructures in the fs-PLD regime have been reported in different conditions~\cite{nanof1,nanof2,np3,gavi}.

Various formation mechanisms have been conjectured for fractal nanostructures in PLD experiments: they could form in flight during the plume expansion~\cite{kask} or by nanoparticle diffusion and aggregation on the surface~\cite{nanof2} or in both ways.
In the literature it was suggested that diffusion of the ablated material after landing should play an important role in the fractal formation.
Such conclusion were based the fact that the fractal dimension of the aggregates~\cite{nanof2} found in PLD experiment is compatible with the fractal dimensions obtained in numerical experiments with diffusion models of nanoparticle aggregation.
On the other side fractal dimension is not a very sensitive quantity: fractals obtained by very different mechanisms can have the same fractal dimension. 
For these reasons different indicators to sort out fractal dynamics
are needed, such as the  area distribution of fractal nanostructures,
which, together with the fractal dimension, can help to discriminate between different formation mechanisms.

Here we analyze  experimental data of Titanium dioxide ($TiO_2$) nanostructures obtained by fs-PLD in air at ambient conditions on different substrates.
$TiO_2$ is a strategic material in many technologically important areas, such as heterogeneous catalysis \cite{27,28,29}, photo-assisted oxidation~\cite{tech4}, optical~\cite{31} and
photovoltaic~\cite{tech7} devices. 
Under the same experimental conditions (laser fluence, pulse duration, polarization, distance of the target from the substrate, see Methods) we found that different nanostructures are formed on different substrates, ranging from single nanoparticles on graphite, ramified fractals on silicon and  long fractal chains on quartz.
This shows that processes occurring on the substrate, after landing of the ablated material, are essential to explain the nanostructures formation. 

We propose a model for fractal formation in which single nanoparticles, formed during the ablation process or during the plume expansion, land on the substrate and then diffuse and aggregate to form larger nanostructures of fractal dimension, see    \ref{mc}. 
We simulate nanoparticle diffusion and aggregation using a Diffusion Limited Cluster Aggregation (DLCA) model~\cite{mea1,DLA1}. 
DLCA models have been extensively used in literature and they can vary in many specific features, such as dependence of diffusion on the nanoparticle  size, degree of  reversible aggregation, mobility of nanoparticles on a larger cluster (edge diffusion). 
Our model is based on a set of minimal assumptions: irreversible sticking and size independent diffusion probability.

The initial nanoparticle area distribution, from which substrate diffusion calculation starts, is retrieved from the experimental data.
To this end we use experimental data from the HOPG substrate (  \ref{mc}a), where nanoparticle aggregation is negligible and the initial distribution is composed mainly of single nanoparticles,    \ref{f1}a.
From this initial nanoparticle area distribution, the DLCA algorithm is able to reproduce the main qualitative and quantitative experimental features for other substrates (silicon, quartz,    \ref{f1}b,c), where nanoparticle aggregation is essential to explain the observed experimental distribution, by changing \emph{only} the simulation time.

The fractal structures obtained numerically look very similar to the experimental ones. 
Moreover our model reproduces not only the fractal dimensions of the nanostructures, but also their area distribution for different densities of the ablated material on the silicon substrate, where most of our experimental data are collected.

\section{Results and Discussion} 

\begin{figure}[!h]
\centering
{\includegraphics[width=12cm, keepaspectratio]{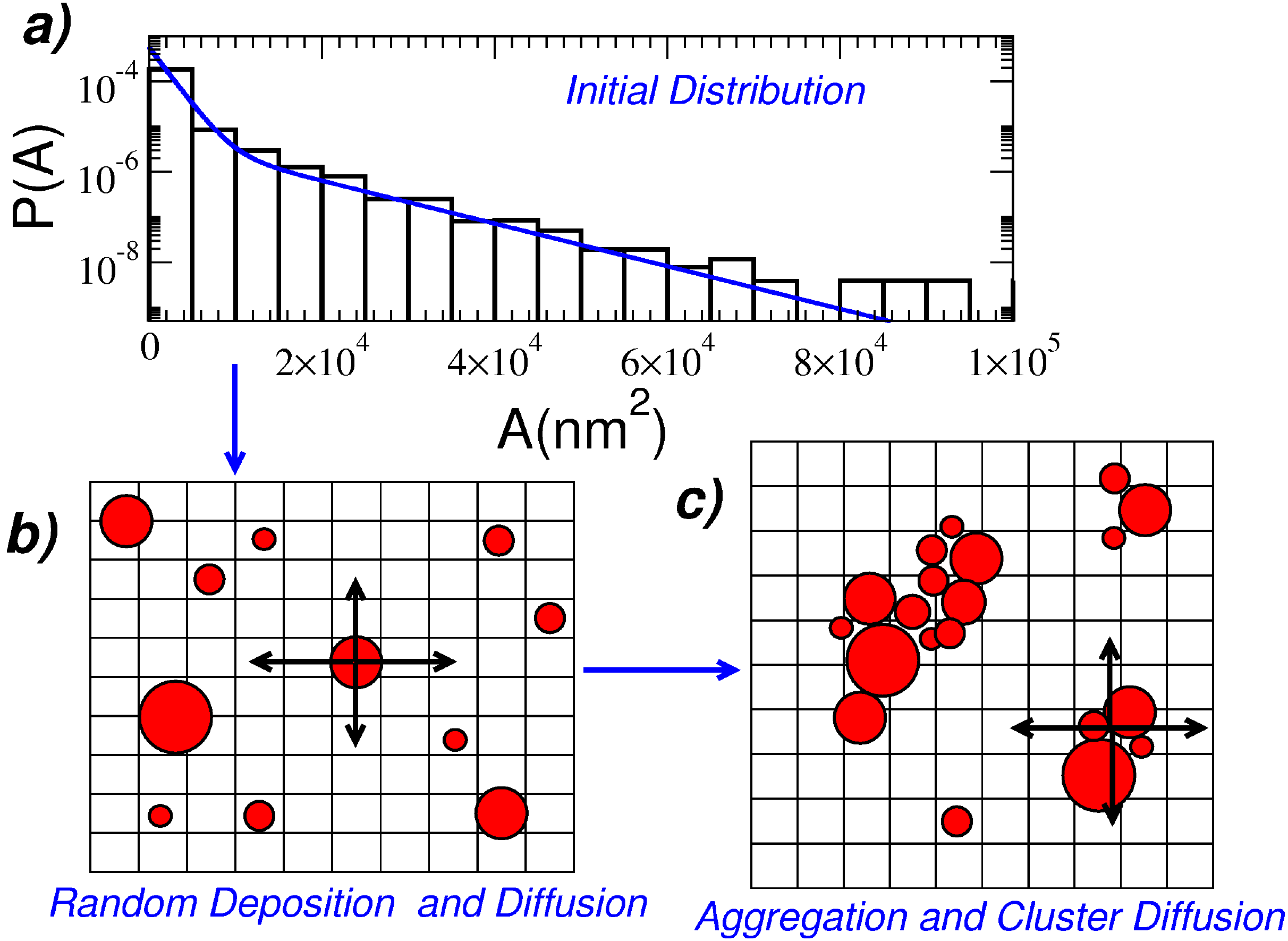}}
\caption{Main steps of the Monte Carlo Method implemented to reproduce fractal structures found experimentally.
a) Histogram of the initial probability density distribution P(A) of
$5 \times 10^3$ single $TiO_2$ nanoparticles areas (A) extracted from experimental data obtained on HOPG substrate, see  (\ref{f1}a).
The histogram is normalized so that $\int_{0}^{\infty}P(A) dA=1$.
The fitting formula (blue curve) is given in  \ref{e1} and normalized as the histogram.
b) spherical nanoparticles of different sizes, extracted from the
probability distribution P(A), are randomly distributed on a two dimensional surface and let diffuse isotropically in the four main directions.
c) when touching nanoparticles aggregate irreversibly.
The cluster thus formed can also diffuse with a size independent probability and aggregate irreversibly when touching.
The process is halted when the same number of clusters per unit area is reached of the experimental images.
For more details see text.}
\label{mc}
\end{figure}

\begin{figure}[!h]
\centering
{\includegraphics[width=12.5cm, keepaspectratio]{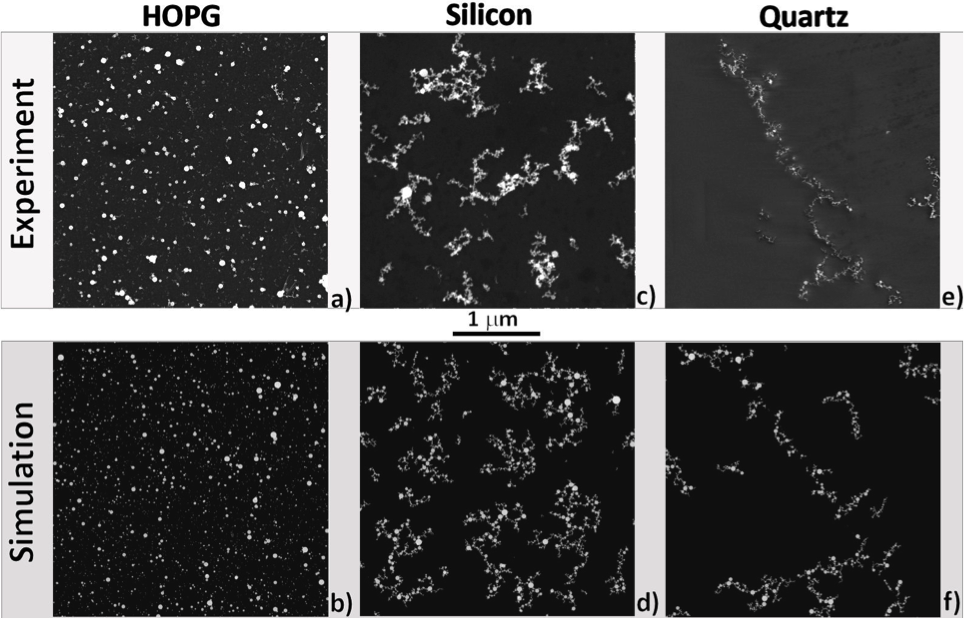}}
\caption{SEM images ($5.6 \times 5.6 \mu m^2$) of $TiO_2$ deposited on
  HOPG (a), Silicon (c) and quartz (e) by fs-PLD, are compared with
  results of numerical simulations. In panels b), d) and f) we show
  the Monte carlo simulations obtained with the diffusive model at the
  same coverage. The numerical simulations shown have been made on a
  grid of $800 \times 800$ pixels and at different coverages, 
corresponding to the experimental images. Transparency were used in
  the simulated images to make them more similar to the experimental
  one.}
\label{f1}
\end{figure}

In \ref{f1}a, c, e, experimental images of $TiO_2$ nanostructures obtained by fs-PLD on different substrates are shown. 
The deposition of $Ti$ by fs-PLD under the same experimental conditions, results in very different nanostructures depending on the substrate types,    \ref{f1}, indicating that
the substrate is playing a main role in determining the aggregation behavior.
The fractal dimension $D_{\rm f}$ of the nanostructures was computed using the counting box method described in Methods.
Results are reported in Table 1.

On HOPG (  \ref{f1}a), we observe no fractal nanostructures, apart from very small aggregates of few NP.
The distribution on HOPG is composed mainly of single nanoparticles
with diameters ranging from $6-7$ nm up to more than $100$ nm, see
histogram plotted in \ref{mc}a. 
On silicon (native silicon oxide,   \ref{f1}c), we observe fractal aggregates randomly ramified, composed of NP with a similar range of diameters.
On quartz we observe an unexpected behavior: structures very similar to one dimensional chains, composed of aggregated NP,   \ref{f1}e, with a fractal dimension (see Table 1).
Such chains are extending up to almost a $mm$ of length with a lateral dimension of few $nm$, with a length/width ratio up to 5 order of magnitude, see   \ref{f2} in Methods. 

\begin{table}
\footnotesize
\centering
\begin{tabular}{|c|c|c|}
\hline
Substrate/Coverage   & Experimental  & Numerical \\
  & Fractal Dimension & Fractal Dimension \\
\cline{1-3}
$TiO_2$ on Silicon &  &  \\
$10\%$ coverage & \normalsize{1.42} & \normalsize{1.36}\\ 
  \ref{f1} c,d) & &\\
\cline{1-3}
 $TiO_2$ on Silicon&  &  \\
 $20\%$ coverage & \normalsize{1.44} & \normalsize{1.45}\\
  \ref{sub1} b,c) & &\\
\cline{1-3}
 $TiO_2$ on graphite &  &  \\
$6\%$ coverage & \normalsize{1} & \normalsize{1}\\
  \ref{f1} a,b) & &\\
\cline{1-3}
 $TiO_2$ on quartz &  &  \\
$5\%$ coverage & \normalsize{1.23} & \normalsize{1.32}\\
  \ref{f1} e,f) & &\\
\hline
\end{tabular}
\caption{Fractal dimension $D_{\rm f}$ of numerical simulations are compared with experimental ones.
For the HOPG, there are no fractals, so that $D_{\rm f}=1$ both for the experimental and the numerical case.
For the case of $TiO_2$ on quartz the $D_{\rm f}$ has been computed for the fractal structures along the defect only.\\}
\label{table}
\end{table}

These data strongly suggest that nanostructures are not formed by aggregation of NP during the plume expansion from the target material to the substrate surface, but they are rather the result of an aggregation mechanism which occurs after landing of the ablated material on the substrate. 
This experimental evidence is the starting point for our model of the NP based structure formation.
We assume that after ablation, NP of different diameters land on the substrate with a random spatial distribution.
The initial distribution of NP diameters for all type of substrates
has been obtained by fitting the experimental area distribution
obtained from the HOPG samples, see  \ref{f1}a, where NP basically do not form larger aggregates and surface aggregation is playing a negligible role.
For the $TiO_2$ NP area distribution, shown in   \ref{mc}a, we obtained the following fitting bi-exponential law: 

\begin{equation}
P(A)=P_1 e^{-A/A_1}+P_2e^{-A/A_2}
\label{e1}
\end{equation}

where $A$ is the NP area in nanometers squared and $P_1=5.58\ 10^{-4}$ nm$^{-2}$, $P_2=5.58\ 10^{-6}$ nm$^{-2}$, $A_1=1700$ nm$^2$, $A_2=9200$ nm$^2$.

After landing on the substrate surface, we further assume that nanoparticles can diffuse and aggregate to form clusters. 
We described such process with a diffusion and aggregation model, belonging to the class of DLCA  models~\cite{mea1,DLA1}, which we implemented using Monte-Carlo simulations.  
The model is based on a set of minimal assumptions, see   \ref{mc}: 
\begin{enumerate}

\item We simulate the impinging ablation plume with spherical NPs arriving on the substrate with a random distribution.
The NP diameter distribution is given by the fitting formula  (\ref{e1}).
The NPs low-diameter cut-off is set to $5 nm$, that is the smallest diameter observed experimentally.
If the initial deposition yields two or more NP touching each other, they are counted as a larger cluster.
At first most clusters are formed by only one NP and the total number number of clusters is counted.

\item  NP and clusters of NP can diffuse isotropically on the substrate surface; each cluster can translate with equal probability in one of the four directions (up-down, left-right) on the surface.
The probability to move in one direction is $p/4$, so that at every time step, each cluster has a probability $p$ to move and $1-p$ not to move.

\item At each Monte Carlo step, the cluster are moved according to their probability and then the number of clusters is counted again.
The clusters of NP formed as a consequence of moving cannot be deformed or rotated, but can only translate on the surface in the four main directions.
When two clusters touch, they stick together irreversibly.

\item We assume that each NP/cluster diffuse with a size/mass independent probability.

\item In order to compare numerical simulations with experimental data, the simulations are halted when the same number of clusters per unit area present in the experimental images is reached.
Note that this choice implies that the numerical results are independent of the precise value of $p$ used in the simulations.
\end{enumerate}

Examples of the numerical results obtained with our model for diffusion and aggregation, are shown in  \ref{f1}b, d, f and can be compared with experimental SEM images shown in the upper panels of  \ref{f1}.
The numerical results show a good qualitative resemblance to the experimental image. 

When TiO$_2$ is deposited by fs-PLD on HOPG, the experimental data in  \ref{f1}a show that no aggregates are formed and the fractal dimension is $D_{\rm f}$=1.
A random deposition of nanoparticles without diffusion, following the probability distribution expressed in  (\ref{e1}), reproduces the experimental data, see  \ref{f1}b.
On the other side, in order to reproduce the ramified fractal structures obtained  on the silicon substrate, we used our model of diffusion and aggregation of nanoparticles. 
The numerical results are shown in  \ref{f1}d for the low coverage case  ($10\%$) and should be compared with the experimental image in  \ref{f1}c. 
The fractal dimension is near the value found from the experimental distribution, see \ref{table} . 
To reproduce the long fractal chains found on quartz substrate, see  \ref{f1}e and  \ref{f2}, we modified our diffusive model further assuming that on the surface there is a defect represented by a line in a given direction.
When a cluster reach such line, diffusion is halted for that cluster, which can however grow if smaller clusters reach it.
The result of numerical simulations within this assumption is shown in  \ref{f1}f.
For this case we chose a line defect with the same direction of the fractal chain in the experimental image, see  \ref{f1}e.

\begin{figure}[!h]
\begin{center}
\subfigure[\tiny{ {\bf SIMULATION: RANDOM DEPOSITION}}]
{\includegraphics[width=5cm, keepaspectratio]{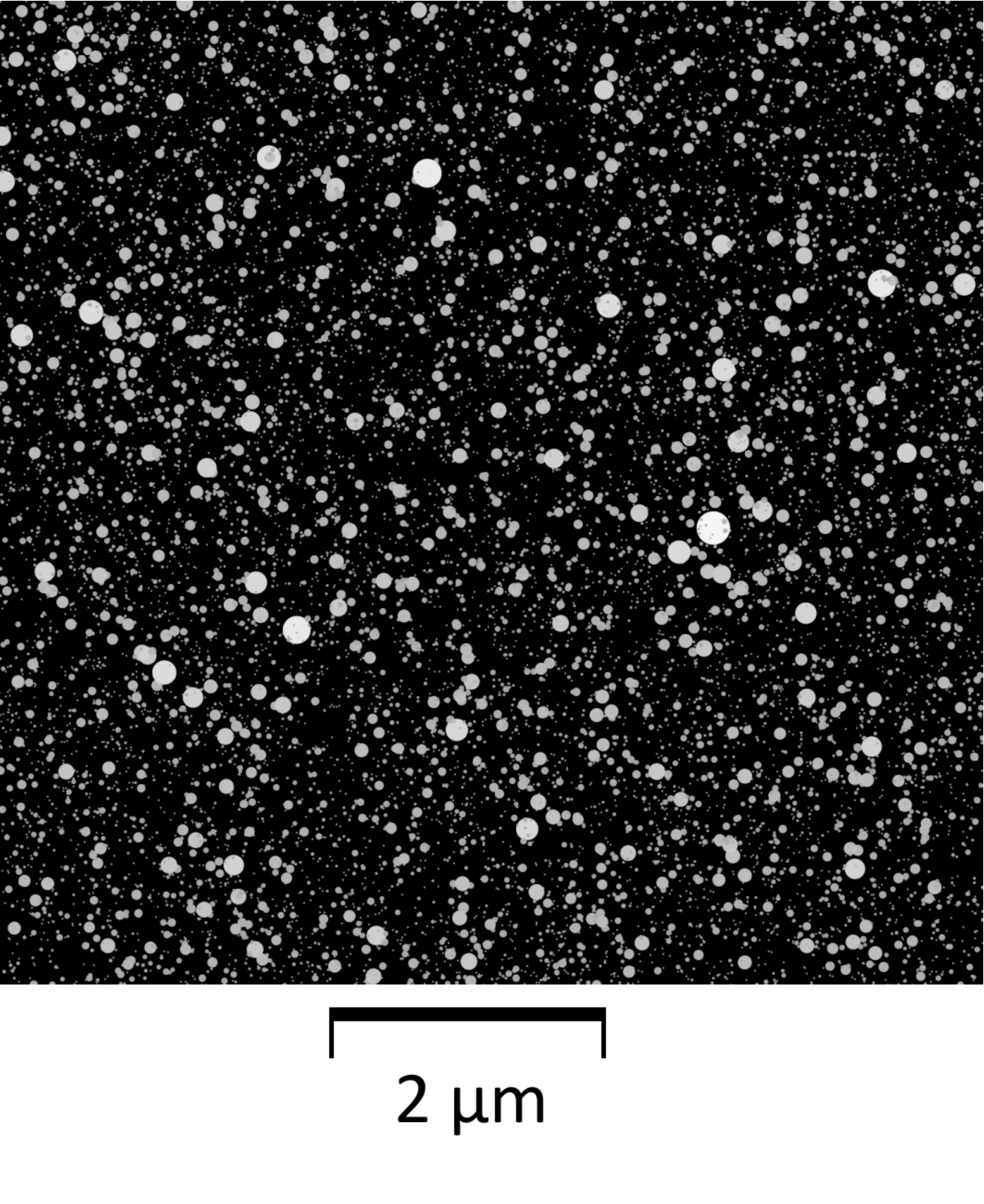}}
\subfigure[\tiny{{\bf SIMULATION: DIFFUSION}}]
{\includegraphics[width=5cm, keepaspectratio]{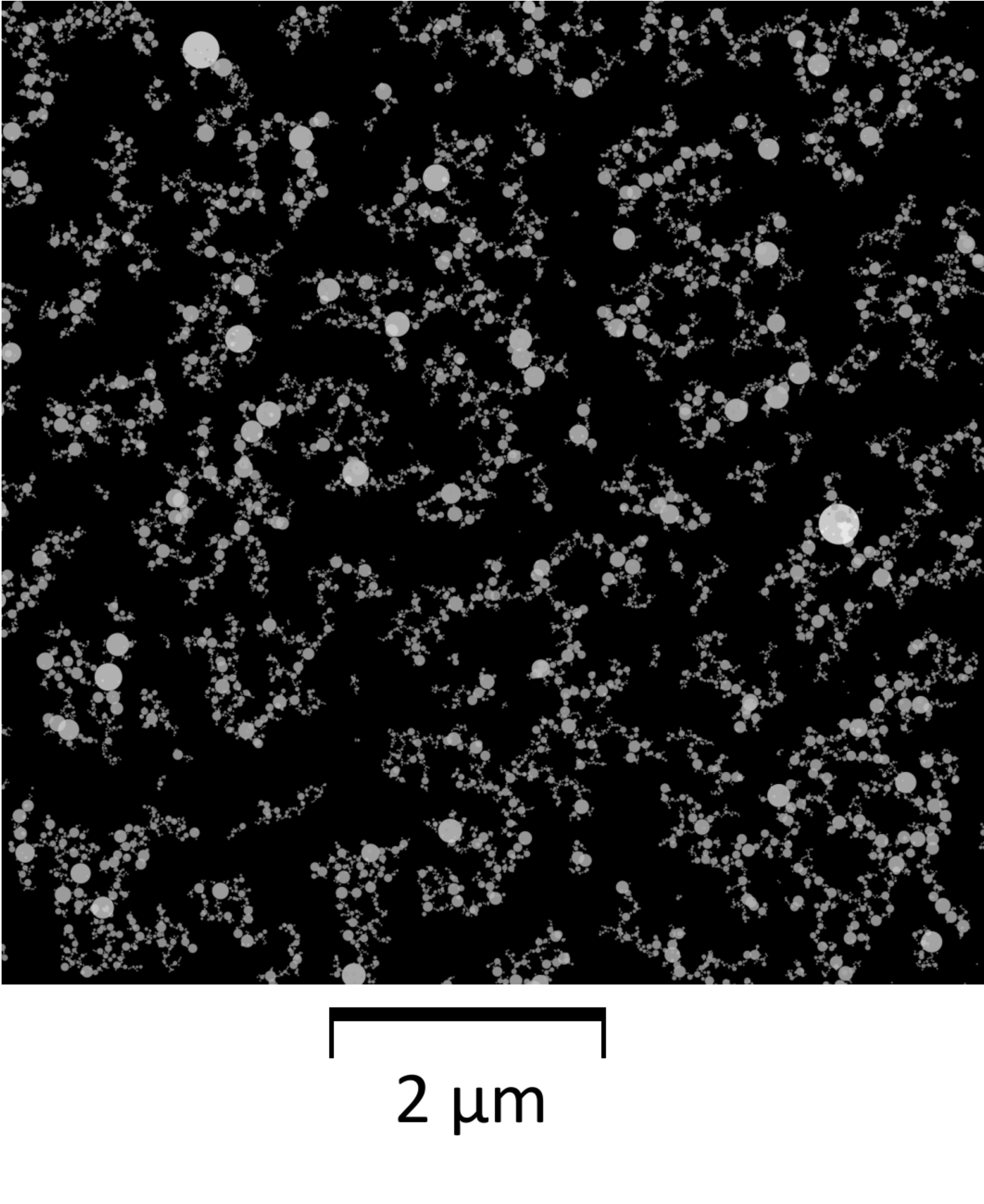}}
\subfigure[\tiny{{\bf EXPERIMENT}}]
{\includegraphics[width=5cm, keepaspectratio]{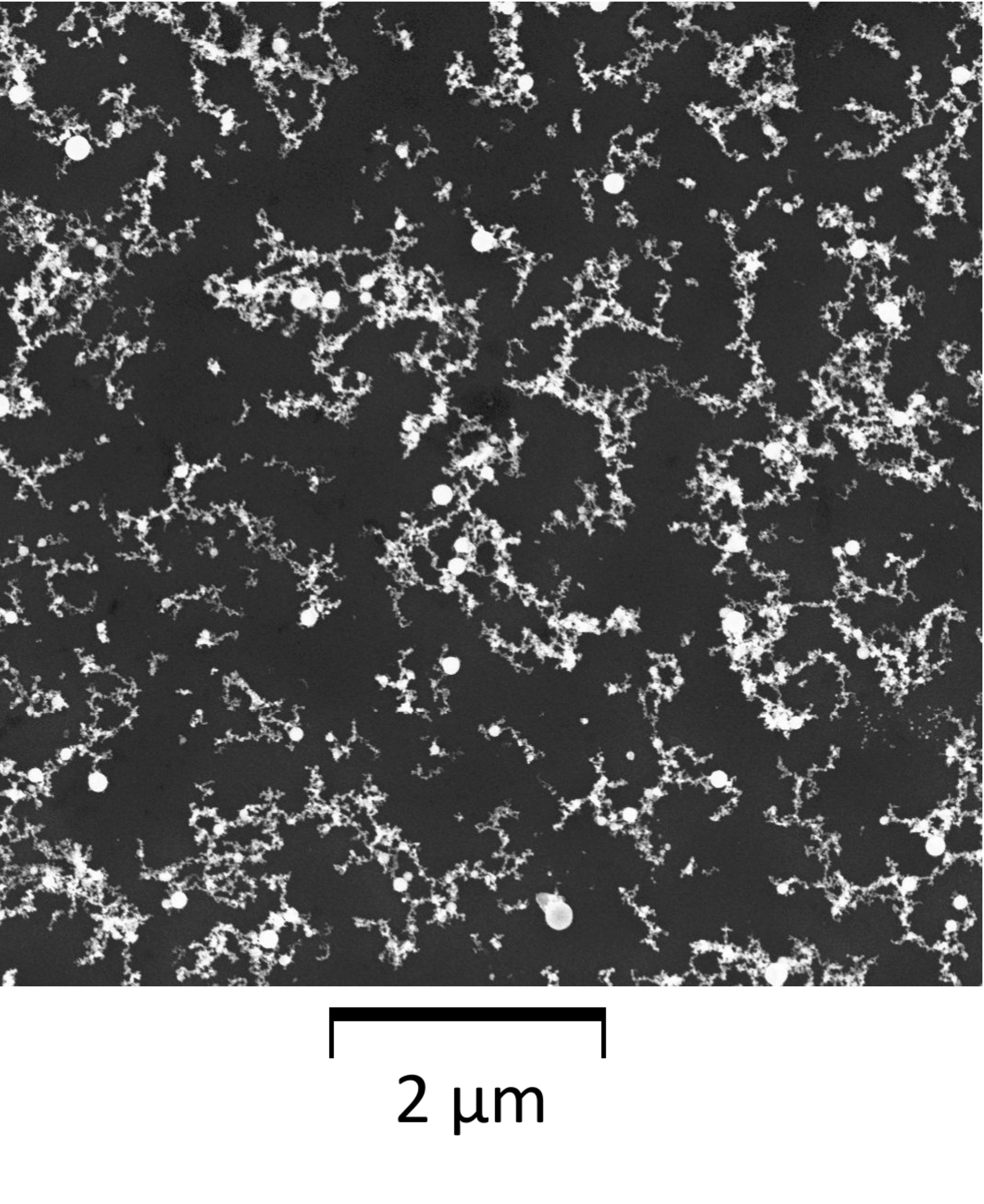}}
\end{center}
\caption{Numerical simulations are compared with experimental images of $TiO_2$ deposited on silicon by fs laser ablation. 
Left panel: Monte Carlo simulation of the initial distribution of nanoparticles without diffusion with the same coverage of the experimental image shown in the right panel.
Central panel: Monte Carlo simulation obtained with the diffusive model described in the text at the same coverage of the experimental image.
Right panel: part of a SEM image of $TiO_{2}$ clusters over a silicon surface with a resolution of $7$ nm per pixel and a coverage of about $20\%$.
The numerical simulations are implemented on a grid of $800 \times
800$ pixels. Transparency were used in
  the simulated images to make them more similar to the experimental
  one.}
\label{sub1}
\end{figure}

In order to highlight the effect of diffusion and to rule out the
possibility that fractal structures could be formed by aggregation of
NP upon randomly landing on the surface, without diffusion, we analyze
in     \ref{sub1} the case of $TiO_2$ fs-PLD on silicon at high coverage ($20\%$).
The result of a numerical simulation where only random deposition of nanoparticles is considered without diffusion and aggregation is shown in  \ref{sub1}a.
The result is clearly different from the experimental data in  \ref{sub1}c, which instead well resembles our numerical results including diffusion in  \ref{sub1}b.
This further confirms  that the NP diffusion is essential to reproduce the experimental results.

\begin{figure}[!h]
\centering
{\Large
\subfigure[{\bf EXPERIMENT}]
{\includegraphics[width=4.25cm, keepaspectratio]{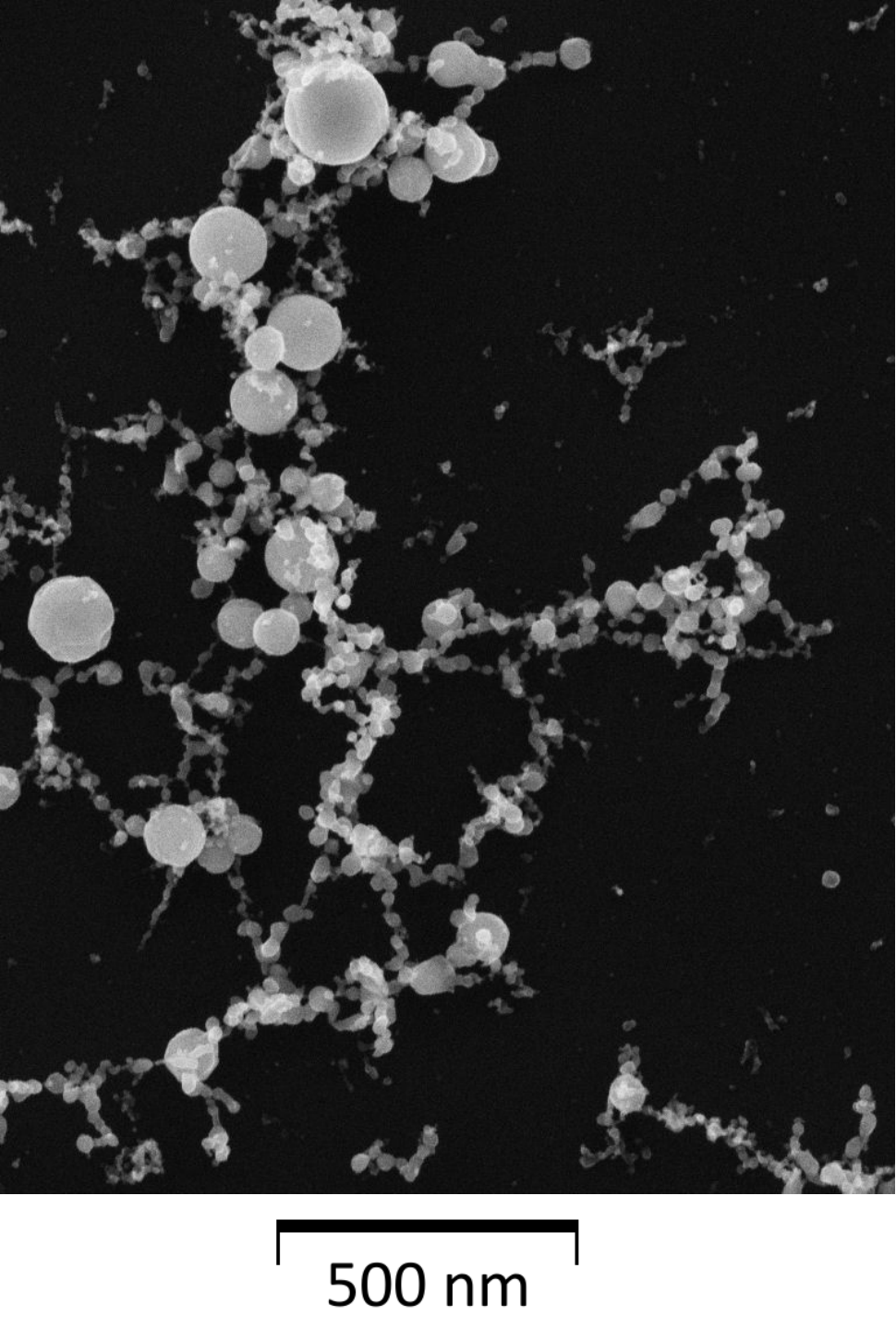}}
\subfigure[{\bf SIMULATION}]
{\includegraphics[width=4.25cm, keepaspectratio]{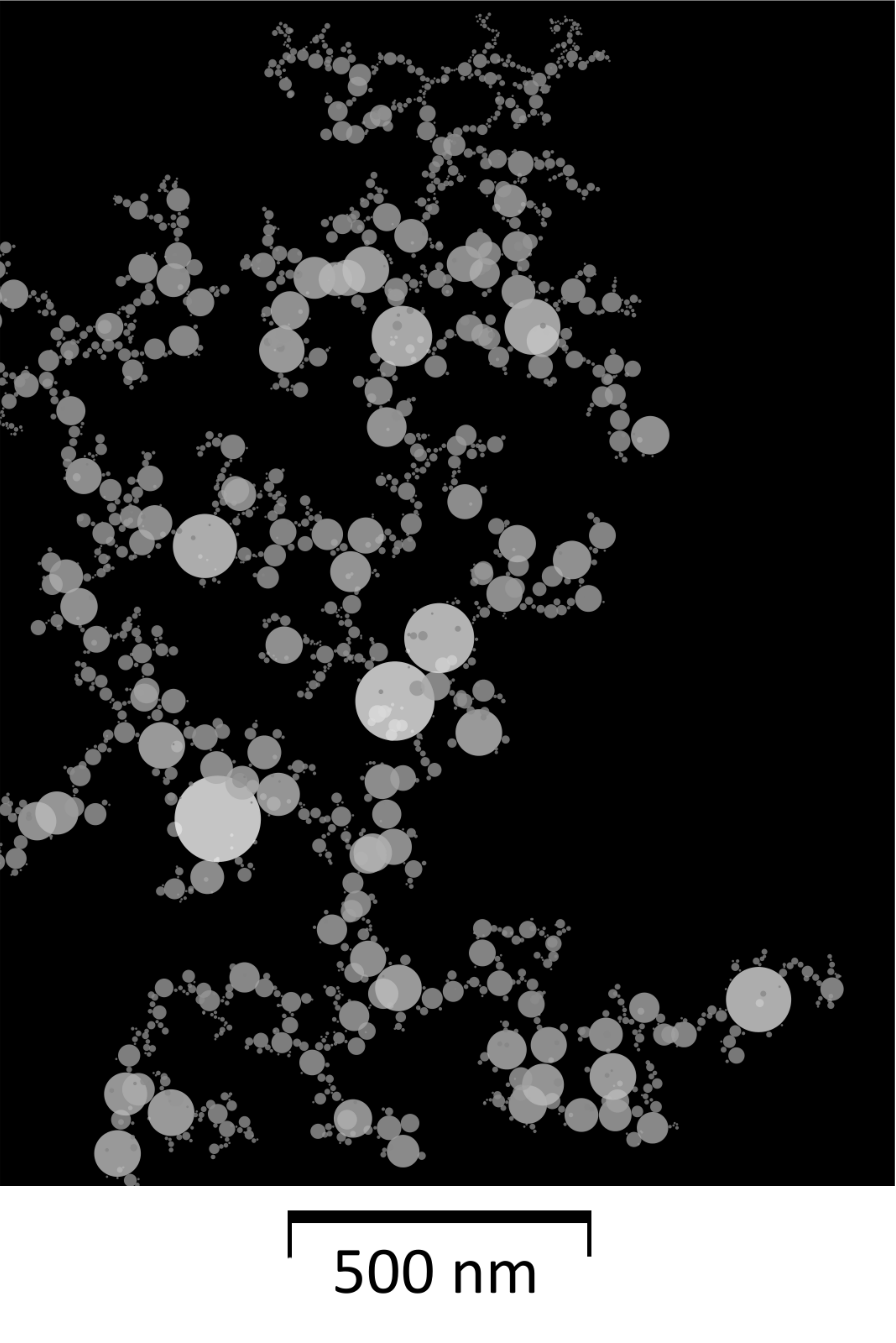}}}
\caption{(a) SEM image of $TiO_{2}$ clusters on silicon surface with a
  resolution of $2$ nm per pixel and about $20\%$ coverage (b)
  simulation made on a grid of $768 \times 1024$ pixels at $20\%$
  coverage. The fractal dimension of the experimental cluster $D_{\rm
    f}\approx 1.36$ is very close to the fractal dimension of the
  nanostructures distribution obtained with our numerical simulation,
  $D_{\rm f}= 1.4$. Note that the NP observed experimentally present
  faceted morphology, indicating a crystalline structure, while the
  simulated NP are circular. Transparency were used in
  the simulated image to make it more similar to the experimental
  one. }
\label{large}
\end{figure}

We have also simulated the case of a single large cluster, see  \ref{large}.
By comparing the experimental image in  \ref{large}a with numerical simulation in  \ref{large}b one can note that the simulation is able to reproduce the aggregation pattern very well down to the smallest NP. 

\begin{figure}[!h]
\centering
{\includegraphics[width=12.5cm, keepaspectratio]{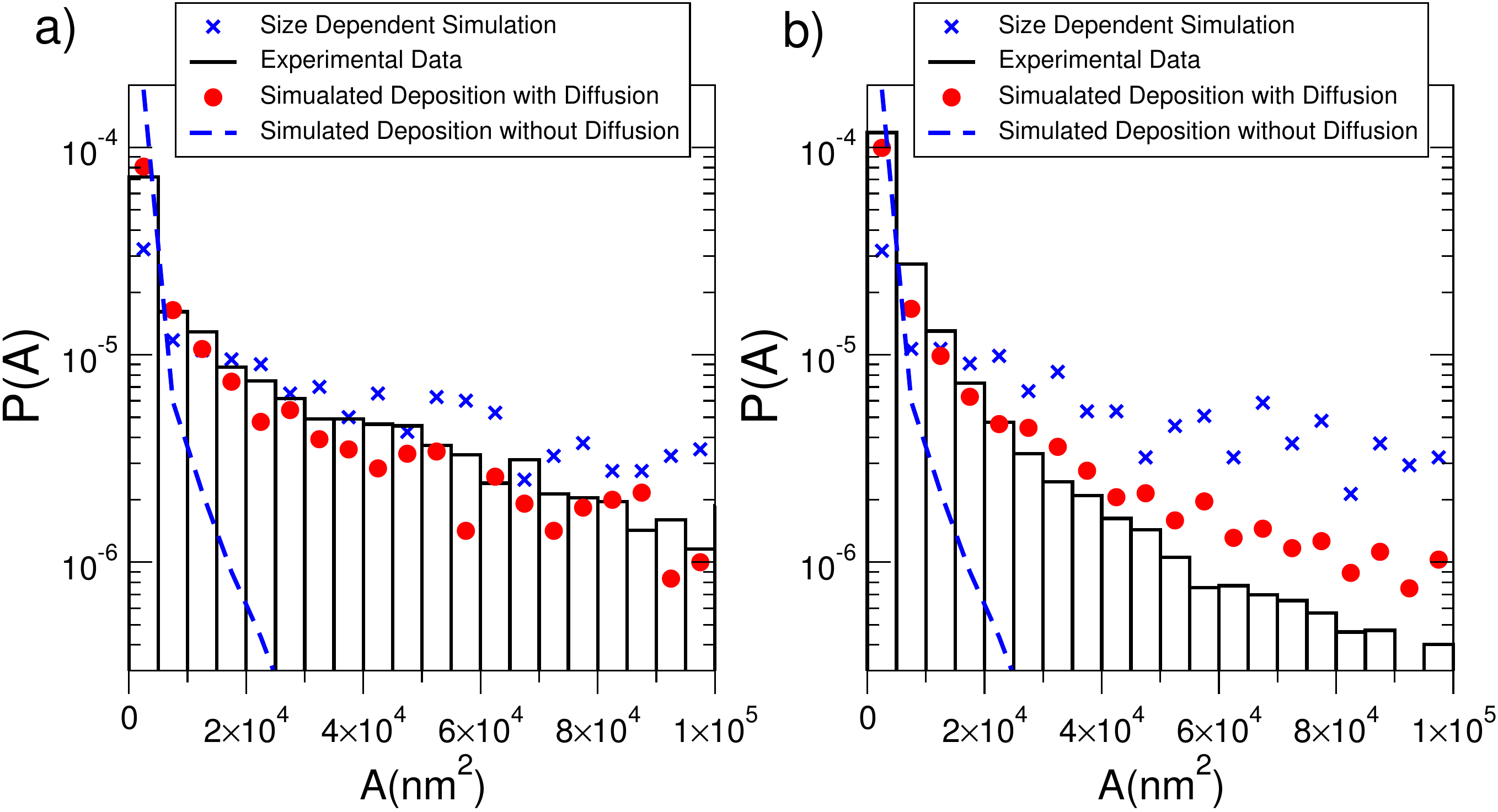}}
\caption{Comparison between the area distributions of fractals.
a) $10\%$ coverage case: black histogram is the area distribution extracted from 2245 clusters deposited on an overall area of $1081 \mu$m$^2$ (7 single images with an area of $155 \mu$m$^2$), while red circles show the result of numerical simulations extracted from 1599 simulated clusters.
b) $20\%$ coverage case: black histogram is the area distribution extracted from $23869$ clusters deposited on an overall area of $6030 \mu$m$^2$ ($19$ single images with an area of $317 \mu$m$^2$), red histogram is extracted from $2998$ simulated clusters.
As a blue dashed curve the distribution of the cluster area obtained just after randomly depositing the nanoparticles on the surface and without diffusion is also shown.
In order to have the same number of clusters per unit area, the Monte Carlo steps for the $20\%$ coverage case were one-third of the steps used in the $10\%$ coverage case. }
\label{a1}
\end{figure}

Note that the fractal dimensions of the simulated nanostructures are in good agreement with the fractal dimension of the experimental images, see   \ref{table}. 
As already stated in the Introduction, different mechanism of fractal formation can lead to structures with similar fractal dimension.
Thus we consider the area distribution of the nanostructures as a further test for the theory.
The experimental data on silicon are a good benchmark since a large statistics is available for the area of the nanostructures for different coverages. 
In  \ref{a1}, the area distributions obtained  for two different coverages on silicon, are compared with the numerical results.
The distribution obtained numerically (red circles) is in good agreement with the experimental distribution, showing the effectiveness of our model. 
For the $20\%$ coverage case, shown in  \ref{a1}b, the algoritm
overestimates the probability for clusters with large area.
This difference could be explained by the fact that
experimentally $150$ laser pulses have been used to
obtain the sample, see Methods. Each pulse is separated by 1 ms, a time probably sufficient for the nanoparticles to thermalize.
In our simulations we neglected the effect of such multiple
depositions, which can become relevant at higher coverages. Indeed for
large coverage newly arriving nanoparticles are more likely to land on pre-existing
clusters, resulting in a smaller area of the final fractal
structures. 

In order to highlight the importance of the diffusive mechanism, we also reported in  \ref{a1} the distribution of the nanostructures obtained just after the random deposition of NP on the surface, before diffusion (blue curve).
The result clearly show that diffusion is essential to reproduce experimental data. 

The assumption of a size independent diffusion might appear counterintuitive, since larger cluster are expected to diffuse less than smaller clusters.
In order to check this assumption, we compared the experimental data with numerical results obtained using a size dependent diffusion probability. 
At variance with the size independent diffusion model, now nanoparticles and clusters of different sizes have different probability to move. 
We implemented a model in which smaller particles are more likely to move. 
Assuming that the mass of clusters is proportional to their area, we
modified the probability to move $p$ as $p \propto \sqrt{
  \frac{A_0}{A}}$, where $A_0$ is the area of the smallest
nanoparticle and $A$ is the area of the cluster we want to move.
As in the previous cases, we halt the simulation when the same number of clusters per unit area of the experimental images is reached.
The fractal dimension obtained with the  size dependent model is in reasonable agreement both with the fractal dimension obtained from experimental data and with the fractal dimension obtained with the size independent model: for the case of $10\%$ coverage we obtain $D_{\rm f} = 1.37$, while for the case of $20\%$ coverage,  we obtained $D_{\rm f} = 1.49$ (see Table 1 for a comparison with experimental data and simulation with the size independent diffusion model).
This confirms that fractal dimension is not very sensitive to the specific mechanism of fractal formation. 
On the other side, the area distribution obtained with the size
dependent diffusion model (see  \ref{a1}), does not agree with
experimental data: the area distribution is more uniform with a lower
number of smaller clusters and a larger number of big clusters with
respect to the experimental data. These features are common to any
size dependent diffusion mechanism. 
Indeed, they can be explained with the fact that large clusters have a 
small probability to move so that growth occurs around many
aggregation centers. 
This result shows that the assumption of the size independent diffusion
reproduces the experimental results in a better way.


{\bf Discussion.}
The data presented here give evidence that fractal nanostructures obtained on different substrates by fs-PLD of $TiO_2$ in air and at room temperature (RT) are formed by diffusion of the deposited NP on the surface.
To rationalize these findings, we propose a possible substrate-dependent mechanism of diffusion.



In PLD fractal structures are formed by nanoparticles with different diameters containing several  thousands atoms~\cite{gavi}.
The ablated material has very high temperature when it arrives on the substrate: the thermal energy is stored in each NP due to the mechanism of plume generation~\cite{fs}. 
After deposition it is reasonable to assume that diffusion continues at a relevant rate until thermalization between NP and substrate, which is at RT, occurs. 
Under the same experimental conditions (laser fluence, distance of the
substrate from the ablated target material, etc..) the NP temperature will be the same on the different substrates.
Thus the properties of the fractal structures will be strongly influenced by the time the NP need to thermalize with the substrate and thus on its thermal conductivity.
We can expect a fast thermalization and small fractals for substrates with large thermal conductivity, while a slow thermalization and the formation of large fractals for small thermal conductivity substrates. This scenario is consistent with the experimental results shown here, see  (\ref{f1}) and in  (\ref{f2}).
For HOPG, which has a large in-plane thermal conductivity ($\sigma_{HOPG}$=1700 W/(m$\cdot$ K)), the ablated NP do not form any cluster.
In the case of quartz, with thermal conductivity ($\sigma_{SiO_2}$=1.4
W/(m$\cdot$ K)) three order of magnitude lower than HOPG, we observe
the formation of chains up to hundreds of $\mu m$ long, see
(\ref{f2}) in supplementary material.
For silicon, which has an intermediate thermal conductivity ($\sigma_{Si}$=70 W/(m$\cdot$ K)), we have the formation of clusters with sizes in between the two preceding cases.
The connection between the properties of fractal aggregates and the thermal conductivity of the substrate is very intriguing and further experimental data and theoretical work are needed to analyze the relevance of the substate thermal conductivity.
The above discussion also suggests that fractal aggregates size should depend on the laser fluence since this parameter will impact the initial temperature of the ablated nanoparticles at landing and might also impact their area distribution. 

The interpretation of our results is based solely on the role of
thermal conductivity. We have neglected the influence of the substrate
roughness and substrate charging (except for the hypotesis on quartz),
for the following reasons: $1)$ the substrate rms roughness  is the lowest for HOPG (less than 0.1 nm), and similar for silicon and quartz (about 0.5 nm rms as deduced from AFM data), but all in the same range. Since the fractals are not present only on HOPG, we can safely hypothesize that it is not playing a role in determining the NP diffusion; $2)$ HOPG is conductive and the silicon we used for our our experiments is doped (resistivity of few tens of $ohm/cm$) hence the NP can transfer their charge to both surfaces after landing. The main difference observed in the fractal morphology is in fact on the quartz surface, and we have provided a possible explanation of the long fractal chains observed in the supplementary materials (\ref{f2}).

Some final clarifications are in order: $i)$ in our simulations, each cluster diffuse with a probability which is independent from its size.
The good agreement obtained with the experimental data, see  (\ref{a1}),  using such assumption could be explained by the fact that larger clusters loose their energy slower than smaller clusters, thus compensating
their smaller mobility;
$ii)$ the diffusive model used in our simulations, the Diffusion Limited Cluster Aggregation (DLCA)~\cite{DLA1},  have been widely used in describing fractal formation in  vapour phase epitaxy \cite{atomfractal,reviewatom}. 
In this case fractal structures are composed of single atoms which are adsorbed on the surface from the gas phase, in vacuum and at very low temperature.
After the deposition, fractal aggregates are obtained by thermal diffusion of single atoms, which move in a random walk until they stick irreversibly to another growing aggregate.
The situation is completely different and more complicated in the case of PLD at ambient conditions where hot nanoparticles containing several  thousands atoms~\cite{gavi} land on a room temperature substrate. 
Despite such complicated  situation,  we have shown that a simple diffusive model is able to give a good description of the experimental results. 

\section{Conclusions}
We analyzed the formation of nanostructures obtained by fs-PLD of $TiO_2$ nanoparticles on different substrates at ambient conditions. 
In summary, the main results are:
\begin{itemize}
\item[a)] Experimental evidence that NP aggregates are formed on the
substrate and not in flight: we report the finding of nanoparticles on HOPG, fractal nanostructures on silicon, and hundreds of microns long fractal chains on quartz.
\item[b)] Explanation of fractal formation mechanism by a simple
  diffusion model with a minimal set of assumptions: size independent
  diffusion and irreversible aggregation. The Monte Carlo simulations
  reproduce very well the observed fractal structures, including their
  fractal dimension and their area distribution. Often only the
  fractal dimension of nanostructures is addressed when comparing experimental data with numerical results.
As shown in this article, the fractal dimension is not very sensitive
to the specific mechanism of fractal formation. Thus, taking into
account other observables, such as the area distribution of the
fractals, is essential when comparing different models of fractal
formation. 
\end{itemize}
The results presented here are rationalized by taking into account the substrate thermal conductivity as major factor in determining the NP aggregation behavior. Hence our results may open new ways for several technological applications to engineer NP aggregates with tailored fractal dimension and area distribution by femtosecond pulsed-laser deposition at ambient conditions.



\begin{acknowledgement}  
G.L.C would like to thank F. Borgonovi for useful discussion. We thanks G.G. Giusteri for useful discussion and suggestions. We thank P. Jacobson, B. Goncalves for useful discussions and comments. 
This research has been partially funded by the Cariplo foundation grant â Controlled nanostructures by low-cost non-thermal laser ablation on metals at atmospheric pressureâ and by Universit$\grave{\rm a}$ Cattolica del Sacro Cuore through D.2.2 grants. 

\end{acknowledgement} 


\begin{suppinfo}

\section{Experimental method}
$TiO_2$ NP were obtained by fs-PLD performed in air at room
temperature (RT) using an amplified Ti:Sapphire laser system. The
laser pulses at a central wavelength of $800$ nm have a time-width of
$120$ fs FWHM at a repetition rate of $1$ KHz, and the chosen laser
fluence is $9.6 J/cm^2$. A fast mechanical shutter was employed to
control the exact number of pulses striking onto the
target. Geometrical parameters such as sample-target distance and
their relative angle are controlled by a home-made manipulator. The
laser beam-target angle is $45^o$, while the relative sample-target
angle is $0^o$ for all data presented. All data have been taken on
samples with ablated material obtained by $150$ consecutive pulses
($0.15$ s deposition time) on target, and then deposited over 
the substrate surface placed at $2$ mm from the ablation target, unless otherwise indicated. 
The substrates used for deposition are standard silicon wafers cleaned
with isopropanol (Chimica Omnia), highly oriented pyrolitic graphite
(HOPG, Micromash) and quartz. SEM images has been obtained with an
Ultra Plus field emission gun (Merlin system from ZEISS) and a 2D-LE electron column
(FEG gun on Gemini II column) at primary beam voltage $E_b = 5 KeV$, using a nitrogen flux for
charge compensation on insulating substrates.

\section{Nanostructures on Quartz}
\begin{figure}[!h]
\centering
{\includegraphics[width=8.5cm, keepaspectratio]{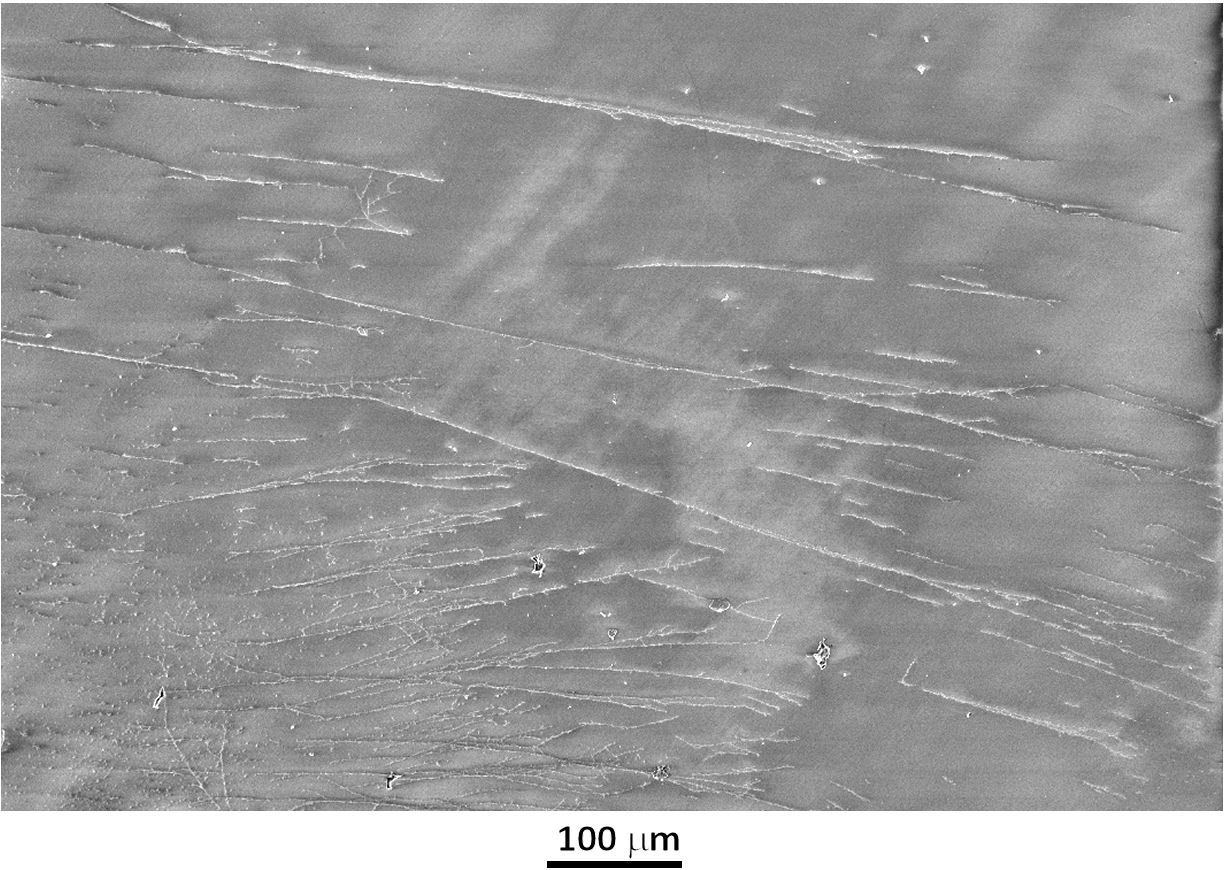}}
\caption{a) SEM image ($870 \times 570 \mu m^2$) of $TiO_2$ deposited
  on quartz in the same experimental conditions as in  (\ref{f1}
  b). 
}
\label{f2}
\end{figure}

 In  (\ref{f2}) chains of nanoparticles extending up to almost a
 $mm$ of length obtained by PLD on quartz are shown.
A possible explanation to the formation of such fractal
chains could be due to the presence of surface defects that are
driving the surface diffusion of the NP along a preferential
direction. The defects might arise from polishing of the quartz
surfac. An alternative
explanation of the formation of one dimensional fractal chains can be
suggested by considering a Coulomb long range interaction between
NP. Since the NP are travelling in air at atmospheric pressure during
the plume expansion, they are interacting with the air molecules, thus
having the possibility of getting charged. When landing on an
insulating substrate such as quartz, the accumulated charge could be
partly maintained during diffusion, giving rise to a long range
repulsion driving the formation of the chains. A similar long range
interaction has been already observed at the atomic scale for alkali
metal deposition on III-V (110) surfaces~\cite{gav2,gav3}.

\section{ Fractal Dimension}
The fractal dimension of the aggregates have been obtained with an homemade
fortran code using  the method of box
counting~\cite{Fractal}: this method consists in counting how many
square boxes, $N(a)$, with side $a$, are needed to cover the contour
of the fractal clusters. The
fractal dimension is then given by the following formula:

\begin{equation}
N(a) \propto a^{-D_{\rm f}}.
\end{equation}

In order to compute the fractal dimension of both the experimental and
numerical images we have isolated  the contour of the fractal
structures. 
For a non fractal curve on a two-dimensional surface we have $D_{\rm f}=1$, since the
number of boxes increases linearly with the size of the boxes, while
for a fractal curve on a two-dimensional surface $1<D_{\rm f}<2$, since the
details of the curve keep increasing by lowering the box sizes. Any real
curve will be fractal only on a certain range of lengths, since below
a length threshold  one cannot resolve fractal structures anymore. 
For this reason for very small $a$ (smaller of the resolution of the
image, or smaller than the typical length scale above which the
structure is fractal) any experimental fractal structure will have $D_{\rm f}=1$.
Also, for very large $a$, the boxes will cover all the surface of the
fractal, thus leading to $D_{\rm f}=2$. Thus the fractal dimension can be
computed in an intermediate range of box-sizes.  



\end{suppinfo}


\begin{thebibliography}{9}
\bibitem{Fractal} 
Mandelbrot B. B., 
The fractal Geometry of Nature, W. H. Freeman and Co., {\bf 1982}.  



\bibitem{acs1} 
 Gottheim S.; Zhang H.; Govorov A.O.; Naomi J.; Halas N.J.
Fractal nanoparticle plasmonics: the Cayley tree. 
{\it ACS Nano} {\bf 2015}, 9, 3284-3292.

\bibitem{acs2} Chen Z.; Pan D.; Zhao B.; Ding G.; Jiao Z.; Wu
  M.; Shek C.-H.; Wu L. C. M.;  Lai J. K. L.
Insight on Fractal Assessment Strategies for Tin Dioxide Thin Films
ACS Nano {\bf 2010}, 4, 1202-1208.



\bibitem{cat} Rigby S. P.; Galdden L. F.
Influence Of Structural Heterogeneity On Selectivity In Fractal Catalyst Structures.
Journal Of Catalysis. {\bf 1998}, 180, 44-50


 
\bibitem{tech8} Gassmann F.;Kotz R.; Wokaun A. 
Supercapacitors boost the fuel cell car.
Europhysics News {\bf 2003} 34, 176-180.



\bibitem{nanof1}  Lin H.; Wiesner M.R.
Deposition of Aggregated Nanoparticles: A Theoretical and Experimental Study on the Effect of Aggregation State on the Affinity between Nanoparticles and a Collector Surface.
  Environ. Sci. Technol. {\bf 2012}, 46, 13270.

\bibitem{nanof2} Nishikawa K.; Takano K.; Miyahara H.;, Kawamura T.;
  Okino A.; Hotta E. 
Nanofractal structure consisting of nanoparticles produced by
ultrashort laser pulse.
App. Phys. Lett. {\bf 2006}, 89, 243112.

\bibitem{28}  Musaev O. R.; Midgley A.E.; Wrobel J.M.; Yan J.; Kruger M.B..
Fractal character of titania nanoparticles formed by laser ablation.
 J. Appl. Phys. {\bf 2009}, 106, 054306.

\bibitem{24}   Ong P.L.; Mahmood S.;  Zhang T.; Lin J.J.; Ramanujan
  R.V.; Lee P.; Rawat R.V.
Synthesis of FeCo nanoparticles by pulsed laser deposition
in a diffusion cloud chamber.
Appl. Surf. Sci. {\bf 2008}, 254, 1909. 

\bibitem{water} Kazakevich P.V.;Simakin A.V.;Voronov V.V.; Shafeev G.A..
Production of copper and brass nanoparticles upon laser ablation in liquids.
 Appl. Surf. Sci. {\bf 252}, 4373 (2006).

\bibitem{np3} Itina T.E.;  Bouflous N.;  Hermann J.;  Delaporte P. 
Nanoparticles and nanostructures formed by laser: what can we learn
from the modeling?
Proc. of SPIE {\bf 2011}, 8414, 841403. 

\bibitem{gavi} Cavaliere E.;Ferrini G.; Pingue P.;Gavioli L.
Fractal TiO2 Nanostructures by Non-thermal Laser Ablation at Ambient Pressure.
J. Phys. Chem. C {\bf 2013}, 117, 23305.  

\bibitem{kask} N. E. Kask N.E.; Leksina E.G.;Michurin S.V.; Fedorov G.M.. 
Fractal nanostructures arising under the ablation of graphite and silicon by millisecond-laser radiation.
Laser Physics {\bf 18}, 762 (2008).  
\bibitem{27}  Amoruso S.; Ausanio G.; Bruzzese R.; Vitiello M.; Wang. X. 
Femtosecond laser pulse irradiation of solid targets as a general route to nanoparticle formation in a vacuum.
Phys. Rev. B {\bf 2005}, 71, 33406.

\bibitem{29}  Koster H.;K. Mann K. 
Influence of beam parameters on the laser induced particle emission from surfaces.
Appl. Surf. Sci. {\bf 1997}, 109, 428-432.


\bibitem{tech4} C. G. Wu C.G.; Chao C.C.; Kuo F.T.
Enhancement of the photo catalytic performance of TiO2 catalysts via transition metal modification. 
Catalysis Today {\bf 2004}, 97,  103-112.

\bibitem{31}  Hastings H.M.; G. Sugihara G.
 Fractals.
 Oxford University Press, New York, {\bf 1993}.

\bibitem{tech7} Que W.X.; Uddin A.; Hu X. 
Thin film TiO 2 electrodes derived by sol-gel process for photovoltaic applications.
J. Power Sources {\bf 2006}, 159, 353-356.


\bibitem{mea1} Meakin P. 
Diffusion-limited aggregation on two-dimensional percolation clusters.
Phys. Rev. B {\bf 1984}, 29, 4327; 
Meakin P.
Formation of Fractal Clusters and Networks by Irreversible Diffusion-Limited Aggregation.
 Phys. Rev. Lett. {\bf 1983}, 51, 1119;
Sorensen C.M.
The Mobility of Fractal Aggregates: A Review.
Aerosol Science and technology, {\bf 2011}, 45, 765-779.


\bibitem{DLA1} Witten T.A.; Sander L.M.
Diffusion-Limited Aggregation, a Kinetic Critical Phenomenon.
 Phys. Rev. Lett. {\bf 1980}, 47,
  1400; Witten T.A.; Sander L.M.
Diffusion-limited aggregation.
 Phys. Rev. B {\bf 1983}, 27, 5686.

\bibitem{fs} Lorazo, P.; Lewis, L. J.; Meunier,
M. Thermodynamic Pathways to Melting, Ablation, and Solidification in
Absorbing Solids Under Pulsed Laser Irradiation. Phys. Rev. B {\bf 2006},
73, 134108;  Perez, D.; Lewis, L. J. Molecular Dynamics
Study of Ablation of Solids Under Femtosecond Laser
Pulses. Phys. Rev. B {\bf 2003}, 67, 184102-184116.


\bibitem{atomfractal}
Brune H.
Microscopic view of epitaxial metal growth: nucleation and
aggregation. Surf. Sci. Rep. 31, 121 (1998);
Brune H.; Romainczyk C.; Roder H.; Kern K.
Mechanism of the transition from fractal to dendritic growth of
surface aggregates. Nature {1994}, 369, 469.

\bibitem{reviewatom} Jensen P.
Growth of nanostructures by cluster deposition : experiments and
simple models. 
Rev. Mod. Phys. {\bf 1999}, 71, 1695.





\bibitem{gav2} 
Gavioli L.; Padovani M.; Spiller E.; Sancrotti M.; Betti M.G.
Self-assembling of potassium nanostructures on InAs(110) surface.
Surface Science {\bf 2003}, 666, 532-535.
\bibitem{gav3} 
 Modesti S.; Falasca A.; Polentarutti M.; Betti M.G.; De Renzi V.;
 Mariani C.
Evolution of one-dimensional Cs chains on InAs (110) as determined by scanning-tunneling microscopy and core-level spectroscopy.
Surface science {\bf 2000}, 447, 133-142.









\end{thebibliography}
\end{document}